\documentclass[preprint,showpacs,showkeys,preprintnumbers,amsmath,amssymb,superscriptaddress]{revtex4}

\usepackage{graphicx}
\usepackage{dcolumn}
\usepackage{bm}

\begin{document}

\def\nuc#1#2{${}^{#1}$#2}
\def\mee{$\langle m_{ee} \rangle$}
\def\mnu{$\langle m_{\nu} \rangle$}
\def\gnu{$\langle g_{\nu,\chi}\rangle$}
\def\mmod{$\| \langle m_{ee} \rangle \|$}
\def\mb{$\langle m_{\beta} \rangle$}
\def\BBz{0$\nu\beta\beta$}
\def\BBm{$\beta\beta(0\nu,\chi)$}
\def\BBt{2$\nu\beta\beta$}
\def\BB{$\beta\beta$}
\def\Gz{$G^{0\nu}$}
\def\Mz{$|M_{0\nu}|$}
\def\Mt{$|M_{2\nu}|$}
\def\Tz{$T^{0\nu}_{1/2}$}
\def\Tt{$T^{2\nu}_{1/2}$}
\def\Tc{$T^{0\nu\,\chi}_{1/2}$}

\title{Multiple-Isotope Comparison for Determining \BBz\ Mechanisms}
\newcommand{\lanl}{Los Alamos National Laboratory, Los Alamos, NM 87545}
\newcommand{\uw}{Center for Experimental Nuclear Physics and Astrophysics, and Department of Physics, University of Washington, Seattle, WA 98195}
\affiliation{	\lanl	}
\affiliation{	\uw	}
\author{	V. M.~Gehman	}\email{vmg@lanl.gov}\affiliation{	\lanl	}\affiliation{	\uw	}
\author{	S. R.~Elliott	}\affiliation{	\lanl	}		

\date{January 12, 2007}
\begin{abstract}
We present a technique for estimating the number of future \BBz\ results using several distinct nuclei to optimize the physics reach of upcoming experiments.  We use presently available matrix element calculations and simulated sets of predicted \BBz\ measured rates in multiple isotopes to estimate the required precision and number of experiments to discern the underlying physics governing the mechanism of the process.  Our results indicate that 3 (4) experimental results with total uncertainty (statistical, systematic, theoretical) of less than  $\sim 20\%$ ($\sim 40\%$)  can elucidate the underlying physics.  If the theoretical ({\it i.e.} matrix element) uncertainty contribution is below $\sim 18\%$,  then 3-4 experimental results of $\sim 20\%$ precision (statistical and systematic) are required. These uncertainty goals can be taken as guidance for the upcoming theoretical and experimental programs. 
\end{abstract}

\maketitle

\section{\label{sec:Intro}Introduction}
\begin{figure}
\includegraphics[angle=0,width=10cm]{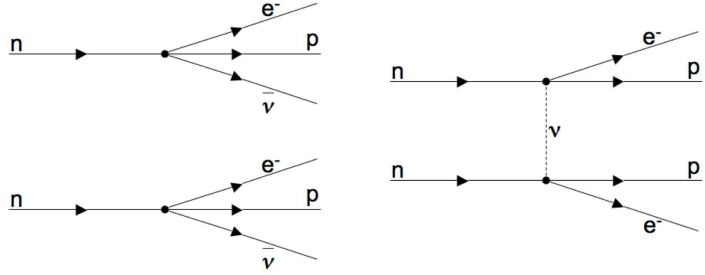}
\caption{\label{fig:LightNuFD} A hadron-level Feynman diagram of \BBt\ (left) and \BBz\ (right) for the light neutrino exchange mechanism.}
\end{figure}

The study of neutrinos in general, and that of double-beta decay (\BB) in particular, is among the most exciting fields on the cutting edge of fundamental physics research as it will guide our development of the standard model of particle physics.  In particular, the need to study \BB\  is well motivated in the literature \cite{Elliott1, Elliott2, Avignone1, Barabash1, Eji05}.   Two-neutrino double-beta decay (\BBt)  ($\it{e.g.}$ $^{76}$Ge $\rightarrow$ $^{76}$Se + 2e$^{-}$ + 2$\bar{\nu}$) is a second-order weak decay process, and is the most rare nuclear decay allowed in the standard model.  \BBt\ can occur in even-even nuclei where beta decay is forbidden, and has been observed in approximately ten nuclei.  Zero-neutrino double-beta decay (\BBz) violates lepton number conservation and is therefore forbidden in the standard model.   \BBz\ differs from \BBt\ by the absence of antineutrinos in the final state ($\it{e.g.}$ $^{76}$Ge $\rightarrow$ $^{76}$Se + 2e$^{-}$).  Though the observation of \BBz\ would imply the existence of massive Majorana neutrinos\cite{SV82} (that is, massive neutrinos that are indistinguishable from their own antiparticle), a number of possibilities exist for the underlying physics mediating the decay.  Figure~\ref{fig:LightNuFD} shows the Feynman diagrams for \BBt\ and \BBz.

This article will focus on the use of a suite of experimental results of the \BBz\ ground-state transitions in different nuclei to distinguish  exchange mechanism models or  transition matrix-element models. This approach is of immediate interest because of the significant number of next-generation experiments that are being proposed that might provide these data. In particular, motivating the required number of experimental results from different nuclei is critical to the overall \BB\ program.

In Section \ref{sec:MECalcOview}, we begin with an overview of the problem of calculating \BBz\ matrix elements, discuss the exchange mechanisms considered in this study, and compile a review of the current matrix element calculations.  Section \ref{sec:PreviousWork} reviews  previous work similar  to that presented here.  Section \ref{sec:XIsoCmp} covers the details of how we performed the model comparison.  Section  \ref{sec:SepAnalysis}  estimates the required total uncertainty as a function of the number of experimental results.  In Section \ref{sec:Conclusions}, we summarize and draw some conclusions about this technique's impact on  the discovery potential of future \BBz\ searches.

\section{\label{sec:MECalcOview}\BBz\ Mechanism and Matrix Element Calculation Overview}
The \BBz\ rate can be written: 

\begin{eqnarray}\label{eq:BBRate}
\Gamma^{0\nu}=G^{0\nu} |M_{0\nu} \eta|^{2}
\end{eqnarray}

\noindent where $\Gamma^{0\nu}$ is the \BBz\ decay rate, \Gz\ is the kinematic phase space factor, \Mz\ is the matrix element corresponding to the \BBz\ transition, and $\eta$ is a lepton number violating parameter (LNVP).  \Gz\ contains the kinematic information about the final state particles, and is exactly calculable to the precision of the input parameters (though use of different nuclear radius values in the scaling factors of \Gz\ and \Mz\ have previously introduced some confusion\cite{Cowell1}).  The LNVP $\eta$ contains all of the information about lepton number violation, and has a form depending on the \BBz\ mechanism.

The LNVP takes on different forms for different \BBz\ mechanisms. In addition, \Mz\ also depends on the mechanism. In this paper we consider: light Majorana neutrino exchange, heavy Majorana neutrino exchange, right-handed currents (RHC), and exchange mechanisms that arise from R-Parity violating supersymmetry (RPV SUSY) models. Our model choices for consideration in this paper were motivated by the existence of calculated \Mz\ for many isotopes within those models. The heavy-particle models represent a large class of theories that involve the exchange of high-mass ($>$1 TeV) particles.  For example, leptoquarks\cite{Hir96a} have very similar \Mz\ to RPV SUSY\cite{Hir96}. Left-right symmetric models can lead to right-handed current models\cite{Doi85} or heavy neutrino exchange models\cite{Hir96b}. Scalar bilinears\cite{Kla03} might also mediate the decay but explicit matrix elements have not been calculated yet. For SUSY and left-right symmetric models, effective field theory\cite{Pre03} has been used to determine the form of the effective hadronic operators from the symmetries of the \BBz-decay operators in a given theory. This last paper makes clear the close connection between all the heavy-particle exchange models. 

	\subsection{\label{sec:MECalc}Matrix Element Calculations}
	The nuclear matrix element in equation \ref{eq:BBRate} contains the information about the nuclear structure of the reaction, and is influenced by the exchange mechanism considered for the transition. Two formalisms have been used to calculate \Mz:  the shell model and the quasiparticle random phase approximation (QRPA).  It is important to note the relative strengths and weaknesses of the QRPA\cite{Rodin1}\cite{Civitarese1}  and shell model approaches\cite{Caurrier2}.  QRPA is capable of including an effectively unlimited number of single-particle states, but a limited number of configurations.  The opposite is true of the shell model.  In this light, it is encouraging that the two methods arrive at similar answers.  Most recent calculations of \Mz\ have been done in the QRPA, renormalized QRPA (RQRPA) or proton-neutron QRPA (pnQRPA) frameworks because the nuclei of interest for double-beta decay are sufficiently large as to make the shell model calculations numerically complex.  Encouraging work, however continues in the shell model, and the confidence in this technique will only  improve as computing speeds and numerical methods become more advanced. Overviews of the matrix element calculations are available in the literature\cite{Fae98,Suh98}.  Recent works by Rodin {\it et al.}\cite{Rodin2} and Suhonen\cite{Suhonen1} provide a written debate concerning the implementation of QRPA. This debate has helped elucidate why the collection of previous QRPA calculations have resulted in numerous differing results. In particular, the free parameters in the theory can be normalized differently.  Additionally, not all calculations have included the same input physics.

\subsection{\label{sec:LNVP}\BBz\ Mechanisms and the Lepton Number Violating Parameter}
In this section we define the LNVP for each of the models. 

\subsubsection{\label{sec:LightNuEx}Light Neutrino Exchange}
If the neutrino is a Majorana particle, it can be exchanged between two neutrons and mediate \BBz. If the neutrino 
is light, the LNVP $\eta_{L\nu}$ has the form:

\begin{eqnarray}\label{eq:LightNuLNVP}
\eta_{L\nu}  = \sum_{k}{U_{e k}^{2}  \xi_{k}  \frac{m_{k}}{m_{e}}},
\end{eqnarray}
		
\noindent where the index $k$ spans the light neutrino states, $U_{e k}$ is the $(e, k)$ element of the neutrino mixing matrix, $\xi_{k}$ is a phase, $m_{k}$ is a neutrino mass eigenstateÕs eigenvalue, and $m_{e}$ is the electron mass.  Detailed discussions of light neutrino exchange \BBz\ and its associated matrix element can be found in \cite{Rodin1, Civitarese1, Caurrier2, Rodin2, Suhonen1, Caurrier1, Simkovic1}.  Further description of the interplay between the LNVP and the neutrino mixing parameters is in \cite{Elliott2}.

\subsubsection{\label{sec:HeavyNuEx}Heavy Neutrino Exchange}
Heavy Majorana neutrinos can also contribute to \BBz.  The LNVP ($\eta_{H\nu}$) however, has a different form:
		
\begin{eqnarray}\label{eq:HevNuLNVP}
\eta_{H\nu} = \sum_{k}{U_{e k}^{2}  \xi_{k}  \frac{m_{p}}{m_{k}}}.
\end{eqnarray}
		
\noindent Here, $k$ runs over the heavy neutrino states, $m_{k}$ is the mass eigenvalue of the heavy neutrino ($m_{k} >>$ 1GeV), and $m_{p}$ is the mass of the proton.  It is important to note that for light neutrino exchange, $\eta_{L\nu}$ is proportional to the light neutrino mass.   In contrast, $\eta_{H\nu}$ is inversely proportional to the heavy neutrino mass.  
Heavy neutrino exchange is discussed at greater length in Ref. \cite{Simkovic1} and references therein.

\subsubsection{\label{sec:RHC}Right-Handed Currents}
In left-right symmetric models\cite{Doi85}, right-handed currents can result in interactions leading to \BBz. Two parameters are used to described this phenomenon. $\eta$ denotes the magnitude of a possible right-handed leptonic current coupling to a left-handed hadronic current in the weak interaction Hamiltonian. $\lambda$ denotes the magnitude of coupling between a  right-handed leptonic and a right-handed hadronic current.  For more information about $\lambda$ and $\eta$ see reference \cite{Mut89}.

\subsubsection{\label{sec:SUSYEx}R Parity Violating Supersymmetry}
\begin{figure}
\includegraphics[angle=0,width=15cm]{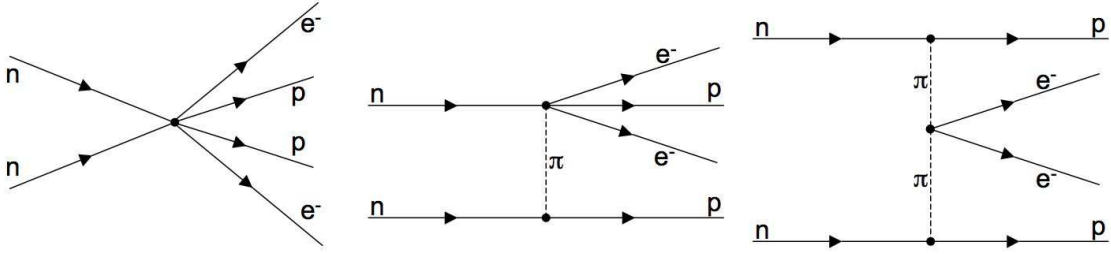}
\caption{\label{fig:SUSYFD} The two-nucleon or short-range contribution (left), and the long-range 1$\pi$ (middle), and 2$\pi$ (right) exchange mechanisms in the heavy-particle exchange framework.}
\end{figure}

Many supersymmetric (SUSY) R-parity violating extensions to the standard model have lepton-number violating interactions between quarks and leptons that can also contribute to \BBz\cite{Faessler1, Wodecki1, Hirsch1, Hirsch2, Hirsch3, Mohapatra1, Babu1, Pas1}.  The particles exchanged are heavy and the physics takes place at short distances. Hence, the heavy degrees of freedom can be integrated out leaving hadron-lepton vertices. The two-nucleon (2N) or short-range contribution to \BBz\ tends to be suppressed because of the large mass of the exchanged particle. Alternatively, the heavy  particle can convert into a virtual $\pi$, where the comparatively low-mass $\pi$ meson can mediate the \BBz\ transition  between more distant nucleons.  If one of the initial quarks gets placed into the virtual $\pi$, we refer to it as the 1$\pi$ mode.  Similarly, if both are placed into a virtual $\pi$, we have the 2$\pi$ mode.  These two modes are discussed in detail in Refs~\cite{Faessler1,Wodecki1}. Generally, many of the heavy particle exchange models have similar \Mz\ with similar short-range and long range components. The direct exchange and its relative suppression are discussed in Refs.~\cite{Faessler1,Wodecki1,Hirsch1, Pre03}. Figure~\ref{fig:SUSYFD} shows the Feynman diagrams for these 3 modes.

For the SUSY \BBz\ mechanisms, the product of $M$ and $\eta_{SUSY}$ take the form\cite{Faessler1}:

\begin{eqnarray}\label{eq:SUSYLNVP}
M \eta_{SUSY} = \eta_{T} M^{2 N}_{\tilde{q}} + (\eta_{PS} - \eta_{T})M^{2 N}_{\tilde{f}} + \frac{3}{8}(\eta_{T} + \frac{5}{3}\eta_{PS})(\frac{4}{3}M^{1\pi} + M^{2\pi})
\end{eqnarray}
		
\noindent Here, $\eta_{T}$ and $\eta_{PS}$ are the effective lepton number violating parameters that normalize the tensor and pseudoscalar currents in the lepton number violating part of the effective Lagrangian.  M$^{2N}_{\tilde{q}}$ and M$^{2N}_{\tilde{f}}$ are the matrix elements for the two-nucleon direct exchange mode, and M$^{1\pi}$ and M$^{2\pi}$ are the matrix elements for the $\pi$ exchange mode.  

\subsection{\label{sec:MEList}Matrix Element Tabulation}

In Table~\ref{tab:PSFac}, we tabulate the phase space integrals in units of $10^{15}$y.  \Gz\ includes a scaling factor ($r_{0}$), which is traditionally factored out and combined with a similar factor from the matrix element to conveniently make the matrix element unitless.  However a problem arises if authors use matrix elements and phase space integrals from different sources and are not careful to make certain that they are scaled by the same value of $r_{0}$\cite{Cowell1}.  In Table~\ref{tab:PSFac} the numerical value of $r_{0}$ used is tabulated for each reference.

\begin{table*}[h]
\begin{center}
\renewcommand{\arraystretch}{0.8}
\begin{tabular}{c|c|c|c|c|c|c|c|c|c|c|c}
\hline\hline
Reference        &r$_{0}$ (fm) &$^{48}$Ca &$^{76}$Ge &$^{82}$Se &$^{96}$Zr &$^{100}$Mo &$^{116}$Cd &$^{128}$Te &$^{130}$Te &$^{136}$Xe &$^{150}$Nd \\ 
\hline
\cite{Simkovic1} &1.1          &          &7.93      &35.2      &73.6      &57.3       &62.3       &2.21       &55.4       &59.1       &269 \\
\cite{Boehm1}   	 &1.2          &75.8      &7.60      &33.5      &69.7      &54.5       &58.9       &2.17       &52.8       &56.3       &249 \\
\hline\hline
\end{tabular}
\caption{\label{tab:PSFac}Phase space factors and the assumed nuclear radius scaling factors. The units are $10^{-15}$/year.}
\end{center}
\end{table*}

Table \ref{tab:LightNuME} contains matrix elements for light neutrino exchange calculated in either the QRPA or shell model framework.  There are literally dozens of other matrix element calculations that are not tabulated here.  As discussed in Refs. \cite{Elliott2, Rodin2}, many of the previous calculations are outdated. We chose the QRPA values for light neutrino exchange in the Table, because we consider them to be the state of the art. We include values for the QRPA from Ref. \cite{Simkovic1,Mut89} for comparison only, since we use values for heavy neutrino exchange and right-handed currents from those references, respectively. For the Shell Model, we used the published values from Ref.~\cite{Caurrier2} except for the cases of $^{100}$Mo and $^{116}$Cd, where we have taken the liberty of using the values from Ref.~\cite{Vogel1} because those values weren't included in Ref.~\cite{Caurrier2}. Since the values in Ref. \cite{Vogel1} have not yet been published, one must consider their use preliminary and only indicative of the anticipated final results.

In Table \ref{tab:HNuSUSYME}, we tabulate the matrix elements arising from heavy neutrino and SUSY particle exchange. The choice of reference for the heavy neutrino exchange mode was made due to its estimates of  \Mz\ for numerous isotopes.  For the SUSY mode, it is the only comprehensive set of calculations done after the realization of the importance of the 1$\pi$ and 2$\pi$ modes. We assumed that the 1$\pi$ and 2$\pi$ terms in Eqn. \ref{eq:SUSYLNVP} dominate, and the values in the table are taken from Ref.\cite{Faessler1} as:

\begin{eqnarray}\label{eq:SUSYM}
M_{SUSY} = (\frac{4}{3}M^{1\pi} + M^{2\pi}).
\end{eqnarray}

\begin{table*}[h]
\begin{center}
\renewcommand{\arraystretch}{0.8}
\begin{tabular}{c|l|c|c|c|c|c|c|c|c|c|c|c}
\hline\hline
Reference          &Method      &r$_0$ (fm) &$^{48}$Ca &$^{76}$Ge &$^{82}$Se &$^{96}$Zr &$^{100}$Mo &$^{116}$Cd &$^{128}$Te &$^{130}$Te &$^{136}$Xe &$^{150}$Nd\\
\hline
\cite{Simkovic1}   &pnQRPA      & 1.1       &          &2.8       &2.64      &1.49      &3.21       &2.05       &2.17       &1.8        &0.66       &3.33 \\
\cite{Rodin2}      &RQRPA\footnote{g$_{A}$ = 1.25, fix coupling parameters to reproduce correct \BBt\ rate}    
                                & 1.1       &          &2.4       &2.12      &0.31      &1.16       &1.43       &1.6        &1.47       &0.98       &2.05 \\
\cite{Civitarese1} &pnQRPA\footnote{fix coupling parameters to reproduce single $\beta$ rate of intermediate nucleus}    
                                & 1.2       &          &3.33      &3.44      &3.55      &2.97       &3.75       &           &3.49       &4.64       &\\
\cite{Mut89}       &pnQRPA\footnote{This paper expresses the matrix element and phase-space factors as a product and doesn't explicity state the r$_0$ value.}
                                &           &          &14.1      &12.3      &          &3.6        &           & 15.2      &9.6        &2.0        &28.9\\
\cite{Caurrier2}   &Shell Model & 1.2       	&0.720     &1.39      &2.19      &          &           &           &0.86       &1.19       &0.75       &\\
\cite{Vogel1}      &Shell Model & 1.2       &          &1.6       &1.7       &0.4       &0.3        &1.9        &2.0        &1.6        &           &\\
\hline\hline
\end{tabular}
\caption{\label{tab:LightNuME}\BBz\ nuclear matrix elements for several light $\nu$ exchange models.}
\end{center}
\end{table*}
		
\begin{table*}[h]
\begin{center}
\renewcommand{\arraystretch}{0.8}
\begin{tabular}{c|l|c|c|c|c|c|c|c|c|c|c|c}
\hline\hline
Reference        &Method & r$_0$ (fm) &$^{48}$Ca &$^{76}$Ge &$^{82}$Se &$^{96}$Zr &$^{100}$Mo &$^{116}$Cd &$^{128}$Te &$^{130}$Te &$^{136}$Xe &$^{150}$Nd \\ 
\hline
\cite{Simkovic1}	 &pnQRPA  & 1.1       &          &32.6      &30.0      &14.7      &29.7       &21.5       &26.6       &23.1       &14.1       &35.6 \\
\cite{Faessler1} &pnRQRPA & 1.2       &          &-625      &-583      &-429      &-750       &-435       &-691       &-627       &-366       &-1054 \\
\hline\hline
\end{tabular}
\caption{\label{tab:HNuSUSYME}\BBz\ nuclear matrix elements for heavy neutrino exchange (upper row) and R-parity violating SUSY models.}
\end{center}
\end{table*}

Table~\ref{tab:RHC} provides the matrix elements arising from right-handed currents as calculated by Ref.~\cite{Mut89}. Again we chose this reference because it provided values for a number of isotopes. However, it does not provide a value for $^{116}$Cd. Therefore, in the analysis described below, these two theories require special treatment.

\begin{table*}[h]
\begin{center}
\renewcommand{\arraystretch}{0.8}
\begin{tabular}{c|l|c|c|c|c|c|c|c|c|c|c}
\hline\hline
Reference    &Method             &$^{48}$Ca &$^{76}$Ge &$^{82}$Se &$^{96}$Zr &$^{100}$Mo &$^{116}$Cd &$^{128}$Te &$^{130}$Te &$^{136}$Xe &$^{150}$Nd\\ 
\hline
\cite{Mut89} &pnQRPA($\eta$)     &          &0.44      &1.54      &          &3.50       &           &0.15       &2.25       &0.83       &29.5\\
\cite{Mut89} &pnRQRPA($\lambda$) &          &0.14      &1.01      &          &1.05       &           &0.007      &1.05       &0.20       &26.8\\
\hline\hline
\end{tabular}
\caption{\label{tab:RHC}Products of \BBz\ nuclear matrix elements and the corresponding phase space factors for right-handed-current interaction models.  The units for the $\eta$ values are 10$^{-8}$/year and for the $\lambda$ values are 10$^{-12}$/year.}
\end{center}
\end{table*}
	
\section{\label{sec:PreviousWork}Summary of Previous Work}
If \BBz\ is observed, it will demonstrate that neutrinos are massive Majorana particles regardless of the underlying physics of the process\cite{SV82}. The prospect of finding a distinctive experimental fingerprint for this underlying physics would then become extremely interesting.  Various techniques have been discussed previously. The use of the kinematic distributions to discern right-handed currents (RHC) from light neutrino mass processes was described in Ref.~\cite{Doi85}. Ref.~\cite{Ali06} discusses the use of the angular distribution as a function of particle physics model. In Ref.~\cite{Tom00}, the author also proposes examining the ratio the rates of \BBz\ to the 2$^{+}$ excited state to that for \BBz\ to the ground state in the same isotope as a signature for RHC. Ref.~\cite{Simkovic2} discusses using relative rates of the decay to the first excited $0^+$ state and the ground state to discern the light neutrino mode, the heavy neutrino mode, and SUSY processes. Ref.~\cite{Cir04} discusses the potential of using $\mu \rightarrow e$ and $\mu \rightarrow e\gamma$ in combination with \BBz\ to prove that the light $\nu$ mechanism is dominant.  Ref.~\cite{Pre06} discusses the relative contributions to $\Gamma^{0\nu}$ from light neutrinos and heavy particle exchange. In this report, we revisit the approach of comparing rates of the ground state transition in multiple nuclei to determine the \BBz\ mechanism.  Singlet neutrinos in extra dimensions can lead to double beta decay\cite{Bha03}. In this case the Kaluza-Klein spectrum of neutrino masses spans values from much less to much greater than the nuclear Fermi momentum. Therefore one cannot separate the nuclear and particle physics and the effective neutrino mass depends on the nuclear matrix element.  The decay rate also depends on unknown parameters such as the brane shift parameter and and the extra dimension radius and therefore is highly model dependent. Finally, mass varying neutrinos (MaVaNs) might lead to decay rates that depend on the density of the matter in which the process occurred\cite{Kap04}.

Three recent articles\cite{Bah04,Bilenky1,Dep06} have considered the question of the required number of \BBz\ experiments.  Reference~\cite{Bah04} took a very pessimistic view of nuclear theory and \BB. That paper took {\em all} previous calculations of \Mz\ and considered them as samples of a parent distribution of the true value. This led to a very wide distribution of possible anticipated results for a given \mee\ and hence a requirement that an impractically large number of measurements be performed.  Many of the calculations, however, have been found to be in error, or have been updated and hence replaced by more recent calculations. (See Ref.~\cite{Rodin2} for a summary discussion.) If one neglects the outlying results, the situation is improved.

Reference~\cite{Bilenky1} concludes that \BBz\ measurements in 3 nuclei would be an important tool in the solution of the nuclear matrix element problem.  This reference does not address the question of discerning the various mechanisms for \BBz, although it does point out that it is an interesting one.  It also does not quantify the confidence one would derive from 3 measurements nor does it quantify the required precision.

During the final preparations of this manuscript, Ref.~\cite{Dep06} became available. This paper estimates the ratio of rates for \BBz\ for a number of mechanisms but doesn't estimate the number or required measurements or their required precision. The paper does estimate the spread in ratios for a number of particle physics models and suggests pairs of isotopes that would be most useful for separating certain models.

\section{\label{sec:XIsoCmp}Multiple-Isotope Comparison}
To compare theoretical calculations and perspective experimental results, we calculate the predicted rate for each considered model using Eqn. \ref{eq:BBRate}.  The LNVP and \Mz\ appear as a product. Hence, if one obtains a decay rate from a lone nucleus, the two factors cannot be individually separated. That is, two different LNVP values can lead to the same decay rate. Therefore, we normalized the LNVPs for each model so that all reproduce identical decay rates for $^{76}$Ge. The absolute normalization was chosen for \mee\ = 100 meV with the matrix elements in Ref.~\cite{Rodin2}; that is, the scale of $\eta$ corresponds to \BBz\ lifetimes just within the reach of the current set of proposed experiments.  This absolute scale for the LNVP is arbitrary for this analysis and is only chosen for definiteness.  Our arguments concerning the uncertainty requirements and the number experimental results do not depend upon it.  

Our model space will consist of seven different theories enumerated below chosen from those described in Section \ref{sec:MEList}.  Because the RHC model does not have a \Mz\ for $^{116}$Cd, our primary analysis uses only the first 5 of these models. A secondary analysis tests the impact of including the RHC models.
  
\begin{enumerate}
\item QRPA($\beta\beta$): Light $\nu$ exchange, RQRPA with coupling constants tuned to reproduce the \BBt\ rate \cite{Rodin2}
\item QRPA($\beta$): Light $\nu$ exchange, pnQRPA with couplings tuned to reproduce the $\beta$-decay rates for the intermediate 
nucleus in the \BBt\ reaction \cite{Civitarese1}
\item SM: Light $\nu$ exchange, shell model \cite{Caurrier2,Vogel1}
\item Heavy $\nu$: Heavy $\nu$ exchange \cite{Simkovic1}
\item SUSY: R-parity violating supersymmetry for the dominant 1$\pi$ and 2$\pi$ exchange modes\cite{Faessler1}
\item RHC-$\eta$: Right-handed current\cite{Mut89}
\item RHC-$\lambda$: Right-handed current\cite{Mut89}
\end{enumerate}

\noindent For each of these models, we simulated 10000 potential sets of decay-rate results for a group of isotopes with {\it measured} decay rates according to a Gaussian distribution whose mean was defined by the predicted decay rate and its variance defined as a fraction of that mean. We then compared each simulated result group to each of the models to discern the predictive power of selecting the correct model. Specifically, each simulated group was compared  to all models by calculating $\chi^{2}$. The best model that fit the simulated set was chosen as that with the lowest $\chi^{2}$ value. We then tabulated the number of simulated sets for which the best choice was also the correct set. We then determined the largest total uncertainty that would result in the correct choice being selected 90\% and 68\% of the time. This analysis was done not only for varying uncertainty, but also for a varying number of isotopes included in the analysis. (Alternatively, one could tabulate the number of times a wrong model was chosen as the correct model and determine how well one can reject a given model.)

The six \BB\ isotopes used in this analysis are: $^{76}$Ge, $^{82}$Se, $^{100}$Mo, $^{116}$Cd, $^{130}$Te, 
and $^{136}$Xe.  These six were chosen because they are proposed for future experiments and have \Mz\ calculations available for a variety of models.  

The analysis presented below is by no means restricted to this set of \BBz\ models and isotopes.  In fact, one would anticipate that the theories will improve and that experimenters will invent new techniques that employ other isotopes. Most importantly, the theory is not entirely settled. Therefore, although we use the $\Gamma^{0\nu}$ values quoted in Table~\ref{tab:ModelValues} as indicative of the difference between the various models, we recognize that the predictions are likely to evolve. It is clear that this work rests on a number of assumptions, including:
\begin{enumerate}
\item We have made the implicit assumption that the present differences in the model predictions are indicative of the true differences. 
\item It is conceivable that future calculations will include additional microphysics (e.g nuclear deformation) that will affect different isotopes differently. Our analysis assumes that any uncertainties in the models shift each \Mz\ similarly. If a model has a systematic shift in its predictions for all isotopes that alters all \Mz\ by an equivalent fraction, that would not alter this analysis in choosing the correct theory, but would result in a shift in the 
value of the deduced LNVP.
\item We have assumed that within each set of theoretical values for \Mz, each calculation for the individual isotopes is of equivalent accuracy.  
\item Finally, we have also assumed that one and only one model is correct. That is, we have assumed that if two mechanisms are contributing to $\Gamma^{0\nu}$ (e.g. light neutrino and heavy particle exchange), one mechanism dominates and any interference is negligible. 
\end{enumerate}
\noindent These assumptions notwithstanding, the analysis can provide useful guidance indicating how precise measurements should be and how many measurements are required.

\section{\label{sec:SepAnalysis}Model Separation Analysis}
\begin{figure}
\includegraphics[angle=0,width=12cm]{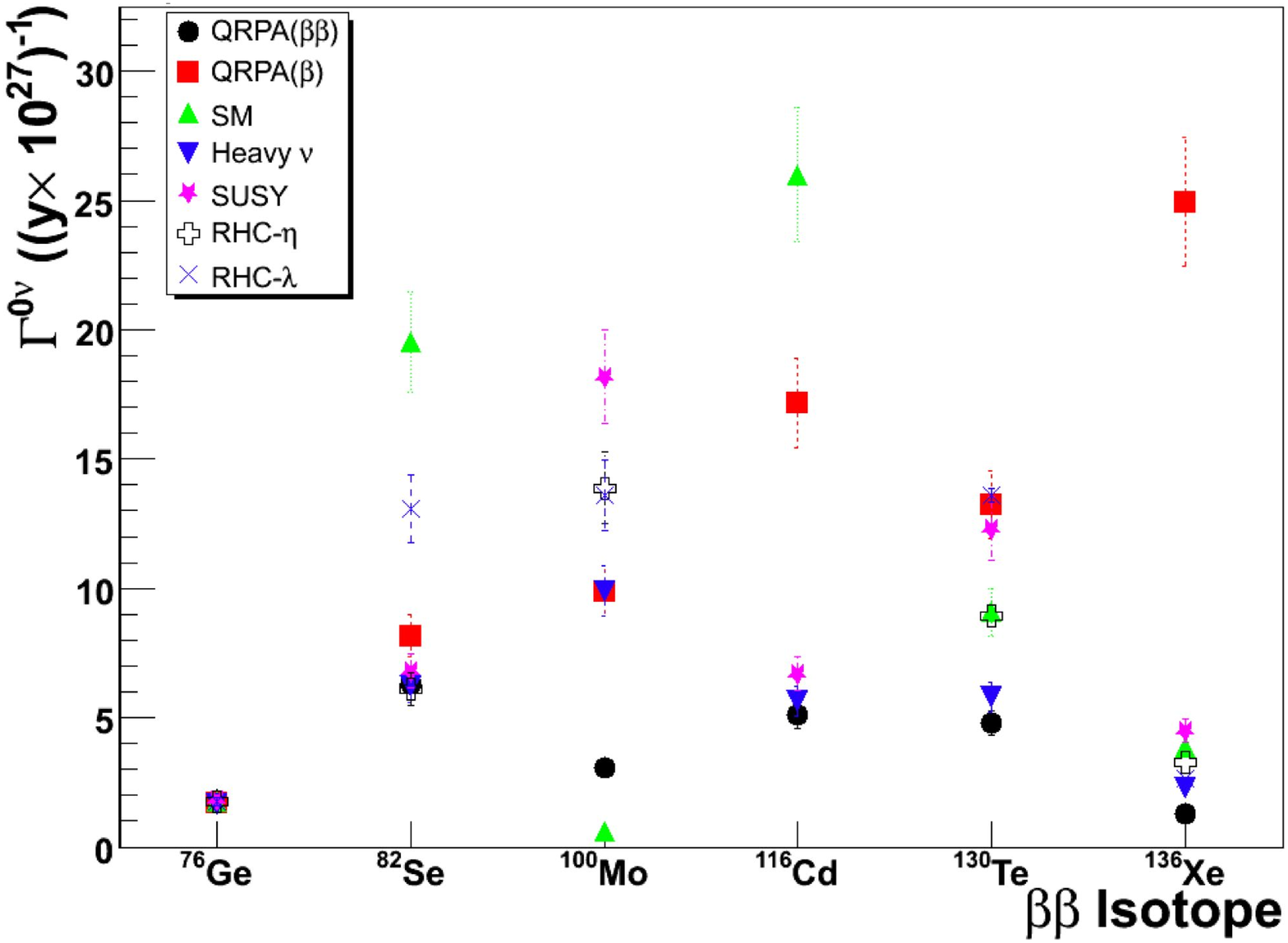}
\caption{\label{fig:ModelPredictions}$\Gamma^{0\nu}$  predictions for \Mz\ uncertainties of 10$\%$ for all 7 \BBz\ models and all 6 isotopes.}
\end{figure}

From the calculated $\Gamma^{0\nu}$ we determine the mean values and uncertainties for each isotope. The explicit values are plotted in Fig.~\ref{fig:ModelPredictions} with 10\% uncertainties shown to provide a qualitative indication of how well the various models can be discerned at this chosen uncertainty level. Values are given in Table \ref{tab:ModelValues}. This table is very useful for identifying which models will be difficult to separate.  For example, the SUSY and heavy neutrino models give very similar decay rates for each isotope to within a somewhat uniform factor of about 2.  Hence it will be hard to separate these two models from each other. Also the table shows the interesting case of the shell model estimate for $^{100}$Mo.  This isotope shows a large disagreement (greater than a factor of 10) with several of the other models.  Hence, this Table indicates that Mo is a key isotope for separating the models. (However note the caveat regarding the reference for the shell model calculations discussed above. This value has not yet been published.) 
	
\begin{table*}[h]
\begin{center}
\renewcommand{\arraystretch}{0.6}
\begin{tabular}{|c|c|c|c|c|c|c|c|}
\hline\hline
Model       	       & LNVP		         &$^{76}$Ge &$^{82}$Se &$^{100}$Mo &$^{116}$Cd &$^{130}$Te &$^{136}$Xe\\
\hline
QRPA($\beta\beta$) &$1.96\times10^{-07}$ &1.75      &6.1       &3.0        &4.9        &4.6        &1.2\\
QRPA($\beta$)      &$1.41\times10^{-07}$ &1.75      & 8.3      &10.0       &17.4       &13.4       &25.3\\
SM                 &$3.38\times10^{-07}$ &1.75      & 19.3     &0.6        &25.7       &9.0        &3.8\\
Heavy $\nu$        &$1.44\times10^{-08}$	&1.75       & 6.6	   &10.5	       &6.0        &6.1        &2.4\\
SUSY               &$7.51\times10^{-10}$	&1.75       & 6.7	   &18.2	       &6.6        &12.3	       &4.5\\
RHC-$\eta$         &$6.30\times10^{-10}$ &1.75      &6.11	   & 13.88     &           &8.92	       &3.28\\
RHC-$\lambda$      &$1.14\times10^{-7}$  &1.75      &13.07     & 13.59     &           &13.59      &2.64\\
\hline\hline
\end{tabular}
\caption{\label{tab:ModelValues}Values of the decay rate in units of 10$^{27}$/year for the isotopes in our analysis for the chosen LNVP. The light neutrino mass (\mee) was chosen to be 100 meV for the model labeled Rodin. The other LNVP's were chosen such that the resultant decay rates for $^{76}$Ge were the same. The form of the LNVP was chosen to be unitless.}
\end{center}
\end{table*}
  
To determine the number of isotopes required to obtain sufficient separation, we varied the number of isotopes included in the  analysis.  By design, the analysis does not include all isotopes, and therefore, the order in which isotopes were added to the analysis had to be chosen.  We chose four isotope orderings; by atomic number, by largest spread in predicted $\Gamma^{0\nu}$, by an assumed likely order of actual results, and an alternate ordering to examine the effect or \nuc{100}{Mo}.  The ``experimental readieness'' (third) ordering and the choice of the isotopes is clearly arbitrary and reflects the opinions, not any prescience, on the part of the authors.  Note: \nuc{76}{Ge} is first in each list is because, as mentioned in Section \ref{sec:XIsoCmp}, we normalized the LNVPs to reproduce the same rate in that isotope.

\begin{enumerate}
\item Atomic number: $^{76}$Ge, $^{82}$Se, $^{100}$Mo, $^{116}$Cd, $^{130}$Te, $^{136}$Xe
\item $\Gamma^{0\nu}$ spread: $^{76}$Ge,  $^{136}$Xe, $^{116}$Cd, $^{100}$Mo, $^{82}$Se, $^{130}$Te
\item Likely order of experimental results: $^{76}$Ge, $^{130}$Te, $^{136}$Xe, $^{100}$Mo, $^{82}$Se, $^{116}$Cd
\item Alternate order to study \nuc{100}{Mo}: $^{76}$Ge, $^{130}$Te, $^{136}$Xe, $^{116}$Cd, $^{82}$Se, $^{100}$Mo
\end{enumerate}

All of the \BB\ isotopes treated in this article have proposals for upcoming Fiorini-style internal source experiments  with the exception of $^{82}$Se. The Majorana\cite{Majorana1} and GERDA\cite{Gerda} collaborations will use $^{76}$Ge. The MOON collaboration\cite{MOON1} plans to use $^{100}$Mo. The COBRA experiment\cite{COBRA1}  will use $^{116}$Cd. CUORE\cite{CUORE1} will use $^{130}$Te and the EXO collaboration\cite{EXO1} will use $^{136}$Xe.  $^{82}$Se is one of the \BB\ isotopes proposed for use by the SuperNEMO collaboration\cite{NEMO1} in a tracking apparatus.

Table \ref{tab:Results} summarizes the results of the simulations.

\begin{table*}[h]
\begin{center}
\renewcommand{\arraystretch}{0.6}
\begin{tabular}{|c|c|c|c|c|c|c|}
\hline\hline
Isotope                          &Confidence &\multicolumn{5}{c|}{ Number of Isotopes}\\
	\cline{3-7}
Ordering                         &Level      &2      &3    &4    &5    &6\\
\hline
Atomic Number                    &90\%       &$<$2\% &8\%  &16\% &23\% &24\% \\
                                 &68\%       &$<$2\% &19\% &36\% &45\% &48\% \\
\hline
$\Gamma^{0\nu}$ Spread           &90\%       &6\%    &18\% &27\% &27\% &24\% \\
                                 &68\%       &13\%   &29\% &41\% &42\% &47\% \\
\hline
Experimental                     &90\%       &3\%    &11\% &24\% &24\% &24\% \\
Readiness                        &68\%       &7\%    &18\% &46\% &47\% &47\% \\
\hline
Alternative                      &90\%       &3\%    &11\% &17\% &15\% &24\% \\
Ordering                         &68\%       &7\%    &18\% &34\% &32\% &47\% \\
\hline
Experimental                     &90\%       &$<2$\% &6\%  &14\% &16\% &\\
Readiness                        &68\%       &$<2$\% &12\% &22\% &24\% &\\
(All 7 models, no \nuc{116}{Cd}) &           &       &     &     &     &\\
\hline\hline
\end{tabular}
\caption{\label{tab:Results}Limits of required total uncertainties corresponding to each theoretical model and the number of measurements as isotopes are added to the analysis for 90\% and 68\% confidence.  Only the final pair of rows includes all 7 models. The others exclude the RHC models because those models do not have a \Mz\ value for $^{116}$Cd. The precision of the numbers in this table are approximately $\pm$1-2\%.}
\end{center}
\end{table*}

\section{\label{sec:Conclusions}Conclusions}
If we examine the results in Section \ref{sec:SepAnalysis}, we see that the ordering of the isotopes has a noticeable impact on the results.  Although, this ordering dependence does indicate that certain models might be better tested with particular isotopes, given the assumptions regarding the nuclear theory upon which these calculations are based, it would be dangerous to put too much weight on the specific choice of isotope for an experiment based on this work using presently available matrix elements.  When considering only 5 models (rows 1-8 in Table \ref{tab:Results}), the results imply that approximately 4 experimental results are required to discern the best physics model underlying the \BBz\ process, depending on the confidence level desired.  For 68\% (90\%) confidence in choosing the correct model, a total uncertainty (theoretical and experimental) of $\sim$40\% ($\sim$20\%) might be acceptable if 4 experimental results are available. In this analysis it is the heavy neutrino model that is the limiting factor in setting the uncertainty levels. If only light neutrino models are considered, then the requirement for 68\% (90\%) can be relaxed to $>$50\% ($\sim$35\%) uncertainty for just 3 experiments. From this analysis, it is also clear that only 2 experimental results is not sufficient. 

When all 7 models are used, the required statistical precision becomes more stringent than when one uses a smaller number of models, as one would expect. A 68\% confidence-level result would require 4 measurements in different isotopes with a precision of $\sim$25\% or better. (See the final pair of rows in Table~\ref{tab:Results}.) The general conclusion remains that separation is possible if the uncertainties can be made small, but it also indicates that additional constraints limiting the possible viable models from non-\BBz\ experimental results will help discern the underlying mechanism. 

The uncertainty described above is the total uncertainty: statistical, systematic and theoretical. The experimental uncertainty (statistical and systematic) for the proposals discussed above can be expected to reach below 20\% or so. This leaves a remaining requirement on the theoretical uncertainly of $\sim$35\% for 68\% confidence-level separation of the 5 models. Because \Mz\ appears squared in the formula for the decay rate, the matrix element theoretical uncertainty is required to be half this value or $<$18\%. This uncertainty requirement on the \BBz\ matrix elements may seem somewhat daunting given their historical spread, but  as discussed in Section \ref{sec:PreviousWork}, there has been tremendous progress in understanding the  source of this  spread.  This is especially true in the case of QRPA-type calculations, providing hope that $10-15\%$ uncertainties are possibile.  In fact, the difference between the shell model and QRPA \Mz\ results are already within $\sim$30-40\% for some isotopes. Since these two approaches are radically different, it seems reasonable to use this difference as an estimate of the precision of these calculations. Such an estimate, however, does not replace the need for uncertainties for the calculations derived from theory itself and from auxiliary measurements\cite{Zub05}.  In any case, should calculations reach these uncertainty levels, it will greatly increase the physics reach of \BBz, and help to formulate the future program of searches for physics beyond the standard model.  

There are other motivations that require multiple \BBz\ experimental results\cite{Ell06}. The need to prove that the observation is indeed \BBz\ and not an unidentified background is not the least among them. Therefore, a general conclusion from this work, and those similar to it, indicate that at least 3 (and very likely 4) \BBz\ experiments along with significant theoretical effort are warranted. The need for and utility of several precision experimental results is the critical conclusion from this work.

\acknowledgments{This work was supported by Los Alamos National LaboratoryÕs Laboratory-Directed Research and Development program. We thank Petr Vogel and John Wilkerson for a careful reading and useful comments.}

\end{document}